 \documentclass[aip,jmp,amsmath,amssymb,reprint,floatfix,onecolumn]{revtex4}
\usepackage{bm} 
\usepackage{amssymb} 
\usepackage{natbib} 
\usepackage{stmaryrd}
\usepackage{graphicx} 
\usepackage{color}
\usepackage{amsmath} 
\usepackage{amsfonts}  
\usepackage{bm}
\usepackage{placeins}
\newcommand{\be}{\begin{equation}}
\newcommand{\ee}{\end{equation}}

\usepackage{color}

\def\REV2#1{{\textcolor{blue}{#1}}}

\def\build#1_#2^#3{\mathrel{\mathop{\kern 0pt#1}\limits_{#2}^{#3}}}

 \newcommand{\vct}[1]{{\mbox {\boldmath $#1$}}}

\begin{document}
 
\title{Deformation statistics of sub-Kolmogorov-scale ellipsoidal neutrally buoyant  drops in isotropic turbulence\footnote{postprint version, accepted for publication on Journal Fluid Mech. vol. 754 pp. 184-207 (2014) }}
 

\author{ L. Biferale$^{1}$, C. Meneveau$^{2}$ and R. Verzicco$^{3}$}

\affiliation{$^{1}$ Department of Physics and INFN, Universit\`{a} di Roma ``Tor Vergata,''  Roma, Italy \\ $^{2}$ Department of Mechanical Engineering, Johns Hopkins University, Baltimore MD 21218, USA \\
$^{3}$ Department of  Industrial Engineering, Universit\`{a} di Roma ``Tor Vergata,''   Roma, Italy \& PoF and MESA+, University of Twente, Enschede, The Netherlands}


\begin{abstract}
Small droplets in turbulent flows can undergo highly variable deformations and orientational dynamics. For neutrally buoyant droplets smaller than the Kolmogorov scale,  the dominant effects from the surrounding turbulent flow  arise through Lagrangian time histories of the velocity gradient tensor. Here we study the evolution of 
representative droplets  using a  model that includes rotation and stretching effects from the surrounding fluid, and restoration effects from surface tension including a constant droplet volume constraint, while assuming that the droplets maintain an ellipsoidal shape. 
The model is combined with Lagrangian time histories of the velocity gradient tensor extracted from Direct Numerical Simulations of turbulence to obtain simulated droplet evolutions. These are used to characterize the size, shape and orientation statistics of small droplets in turbulence. A critical capillary number is identified associated with unbounded growth of one or two of the droplet's semi-axes.  Exploiting analogies with dynamics of polymers in turbulence, the critical capillary number can be predicted based on the large deviation theory for the largest Finite Time Lyapunov exponent quantifying the chaotic separation of particle trajectories. Also, for sub-critical capillary numbers near the critical value, the theory enables predictions of the slope of the power-law tails of droplet size distributions in turbulence. For cases when the viscosities of droplet and outer fluid differ in a way that enables  vorticity to decorrelate the shape from the straining directions, the large deviation  formalism based on the stretching properties of the velocity gradient tensor loses validity and its predictions fail.  Even considering the limitations of the assumed ellipsoidal droplet shape, the results highlight the complex coupling between droplet deformation, orientation and the local fluid velocity gradient tensor to be expected when small viscous drops interact with turbulent flows.  The results also underscore the usefulness of  large deviation theory to model these highly complex couplings and fluctuations in turbulence that result from time integrated effects of fluid deformations.
\noindent
{\bf Key words: breakup/coalescence, drops and bubbles, homogeneous turbulence}
\end{abstract}
 
\maketitle 
 
 \section{Introduction}
 
Improving understanding and characterization of drop deformations and possible breakup in turbulent flows is relevant to a wide range of applications, including engineering processes such as emulsification, homogenization, mixing, blending and multiphase chemical reactions \citep{Davies85,Lefebvre89,Sundaresan00}.  Transport and mixing processes occurring during oil spills and design of remediation strategies also depend critically upon knowledge of oil droplet dynamics and breakup processes occurring in the ocean \citep{ligarret98,Yangetal14}.

Models for the breakup process are dependent upon characterization of the droplet deformations that precede and facilitate breakup.  Much work has focused on breakup and deformations in turbulence when drops are larger than the Kolmogorov dissipation length.  The phenomenological model proposed by \cite{Kolm49} and  \cite{Hinze55} 
 focuses on distorting turbulent stresses as function of scale in the inertial range and compares these with the restoring forces owing to surface tension. This phenomenological  model forms the basis of most of the current models 
for predicting drop breakup in turbulent flows. Numerical simulations based on fully resolved fluid-fluid interface have also been recently performed to study such complex situation \citep{Perlekar12}.
The paper by  \cite{Lasheras02} reviews the salient aspects of large droplets subjected to inertial forces.  
Even without breakup, the understanding of droplet deformation is important in various applications such as predicting effective rheological properties of suspensions or understanding the behaviour of red blood cells interacting with flows including localized large shearing regions.  
In the latter context several studies have been carried out, relying on an analogy between red blood cells and drops, in order to quantify the hemolysis phenomenon \citep{Arora06,detullio12}.

Deformation of droplets may also be due to purely viscous shear forces rather than inertial ones. 
This is of particular importance for droplet-laden turbulent flows when droplets are smaller than the 
Kolmogorov scale. At such scales deformations arise due to viscous drag associated with the shear in the surrounding  flow being resisted by surface tension effects. 
First analyses of droplets in simple viscous shear flow were performed in \cite{Taylor32}. For particular laminar shearing flow, droplets achieve  elongated equilibrium shapes. If the shear is strong enough, a  droplet  may continue to deform and the resistance to deformation due to surface tension is insufficient, leading to unbounded growth
of one or two of the droplet's semi-axes.  This then provides a possible condition for breakup of droplets when subjected to a simple laminar shear flow. The dimensionless number comparing viscous and surface tension  forces is the capillary number $Ca = {\mu_o R G}/{\Lambda}$ ($\mu_o$ is the surrounding fluid viscosity, $R$ a droplet characteristic scale, $G$ is the shear rate, an inverse time scale, and $\Lambda$ the surface tension parameter). The capillary number can be used to characterize the critical conditions (a critical capillary number, $Ca_{c}$), under which stable stationary droplets are no longer possible and  hydrodynamic instabilities develop followed by eventual droplet breakup. 

In laminar flow, the external fluid shear can be characterized by  one or a few parameters associated with the velocity gradient tensor. Conversely, in a turbulent flow, droplets are subjected to a wide distribution of shear/strain rates. In particular due to inner intermittency,  as the Reynolds number grows so does the range of 
values of the local strain and/or shearing rates.  Locally, these can achieve values that exceed the mean value by a 
several orders of magnitude. 
Therefore, one expects that locally some fraction of the droplets will encounter shear rates such that instability and unbounded elongation results. Clearly one would wish to characterize the resulting droplet dynamics statistically. For example, one is interested in probability distribution functions (PDF) of the characteristic scales (e.g. semi axes in the case of ellipsoidal droplets), or orientation dynamics of the droplets with respect of the turbulent flows. If some fraction of the droplets is subjected to unbounded elongation, 
can statistical descriptions still be formulated?  How to characterize statistical distributions of small deforming droplets in turbulence remains an interesting challenge that has not received sufficient attention.  

\begin{figure}
\centering
\includegraphics[width=10cm]{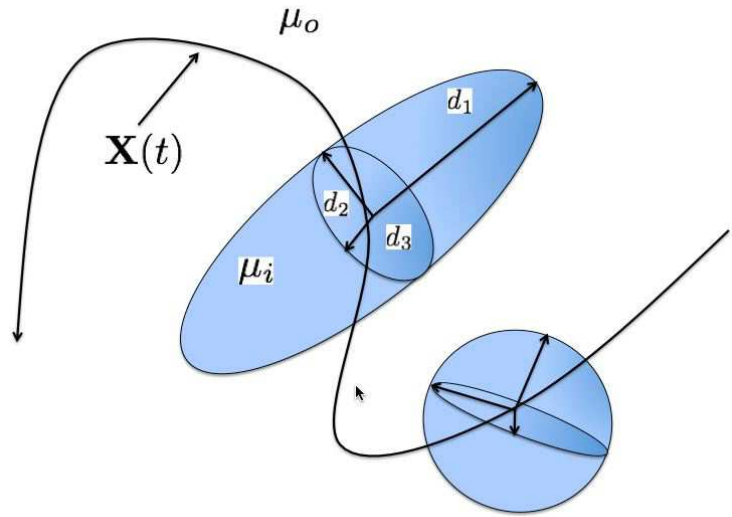}
\caption{\label{droplet}  Typical ellipsoidal shape of one droplet along the turbulent trajectory. The droplet centre of mass ${\bf X}(t)$ is supposed
to follow the evolution of a fluid particle and the droplet deformation is governed by the statistics of the fluid velocity gradients along the 
trajectory, $\partial_i u_j({\bf X(t)},t)$. We denote by $\mu_o$ the viscosity of the carrying fluid and by $\mu_i$ that inside the droplet. 
The droplet is always assumed of ellipsoidal shape with the three 
semi--axes ordered as $d_1 \ge d_2 \ge d_3 $.}
\end{figure}

\cite{Cristinietal03} considered the case of droplets that are smaller than the Kolmogorov scale.
Detailed calculations that combined simulations of turbulent flows
at moderate Reynolds numbers with $Re_{\lambda} \sim 50-60$ ($\lambda$ being the turbulence Taylor microscale) were coupled to a refined boundary integral simulation of local drop dynamics. Such modeling of the droplets
was capable of reproducing highly complex shapes such as necks, their instabilities and precursors of satellite droplet formation. 
Further work along these directions include \cite{Trygvasson09} and \cite{Prosperetti12}.
However, the turbulent Reynolds numbers that can be considered for such highly detailed simulations are relatively limited. It becomes of interest to seek appropriate simplifications that  enable to explore a broad range of turbulent fluctuations of the shear rates to which small droplets can be expected to be subjected in a turbulent flow
at more elevated Reynolds numbers. 
Assuming that the initial shape of the droplets is spherical, the initial deformations lead to ellipsoidal shapes. 
Being characterized by three major axes and their orientations, an ellipsoidal drop shape is much easier to describe and parameterize.  The fate of deforming ellipsoids in turbulence raises a number of interesting questions such as: Denoting the `size' of droplets as the scale of its largest 
semi-axis $d_1$ (see figure \ref{droplet}), we may ask what is the resulting PDF of  $d_1$ in turbulence as function of Capillary and Reynolds numbers? 
Under what conditions can equilibrium distribution functions be found? 
What are characteristics aspect ratios among largest and smallest semi-axes? Do ellipsoidal droplets tend to be axisymmetric or triaxial? 
What are the orientation statistics of deforming droplets? Do their major axes tend to align with the vorticity (as happens with rigid elongated particles \citep{Parsaetal12,ChevillardMeneveau13}), do they align with the most extensive strain-rate direction, or with some other direction? How do such orientation trends
depend on capillary and Reynolds numbers? 
How are these alignments related to the deformation rates? These are some of the questions we will address in this paper. 

Many models are available for the droplet evolution in turbulent flows. 
As a first step, it is important to clarify if one is interested in small or large 
droplets (with respect to the Kolmogorov scale), if they are passive tracers or have 
a feedback on the flow, if the collision/aggregation of different droplets can be neglected
(large dilution limit), if they have inertia (i.e. their centre of mass 
follows or diverges from a fluid particle trajectory). Finally one also needs to specify the 
flow properties of the carrier fluid. In this study we apply several simplifying assumptions
in order to be able to exploit state--of--the--art numerical simulations of turbulent flows 
and  to track many droplets simultaneously.  We focus on  the simple (but still interesting) 
case of sub--Kolmogorov--scale inertialess droplets, in the highly diluted case, therefore neglecting 
the feedback on the flow and the interactions between different droplets. Moreover, we 
assume that the droplet shape can be parameterized by an ellipsoid. Even considering 
this basic case, different models can be adopted. Among them, a popular one has been proposed 
by \cite{MaffettoneMinale98} (hereafter referred to as the `M\&M model'). This model is based 
on the idea that the droplet deformation is the result of the balance between the local stretching terms 
of the velocity gradient  and the restoring surface tension force. An extra non--linear constraint is added 
in order to enforce the preservation of the droplet volume during its evolution. 
The above dynamics can be parameterized by introducing two functions which depend only on the viscosity ratio 
between the droplet and the surrounding fluid.

In the present work we examine the fate of small ellipsoidal droplets being transported and 
distorted by homogeneous isotropic turbulence following the M\&M model. We study different statistical 
properties of the droplets' shape and its correlation with the underlying turbulent 
fluctuations when changing the viscosity ratio. It is worthwhile noticing that, when the 
surrounding fluid and the droplet inner fluid have the same viscosity, the M\&M model is very similar to studying the 
advection/stretching of a small fluid volume together with a relaxation towards a spherical 
shape. This is also the set--up describing the evolution of the second--order conformation 
tensor of simple passive polymers in the Oldroyd--B model. In both cases, the deformation rate 
can be predicted in terms of the statistics of the Lyapunov exponents  governing the chaotic 
properties of particle trajectories; we will exploit this similarity in order to predict the 
critical Capillary number $Ca_c$ where all droplets will break with probability one for such viscous ratio.  

However, when the viscosity ratio of the fluids is strongly differs from unity, the
above analogy does not hold and the prediction of the critical capillary number fails. 
While many typical applications have viscosity ratios different from unity (e.g. oil droplets in water), the
present analysis for unity viscosity ratio is still of interest since there are  
examples of real liquids with viscosity ratios equal to or close to one that are relevant in technology or Nature.
One of them is the pair polydimethylsiloxane/polyisobutylene that are both Newtonian fluids, widely
used as lubricant and in a large variety of other applications. Their dynamic
viscosities at $23\ ^o$C are, respectively, $103$~Pa\ s and $101$~Pa\ s
and their interfacial tension is $2.4$~mN/m \citep{Guido2000}.
Another relevant context is the red blood cells in the plasma matrix:
in this case the viscosity ratio between the hemoglobin inside the
cells and the plasma is less well defined since it depends on the link
with $O_2$ or $CO_2$ of the hemoglobin molecules and on the inner cell
cytoskeleton consisting of proteins. Nevertheless a viscosity ratio in
the range 3--5 is commonly adopted as a reasonable parameter for a
healthy human being \citep{Pozrikidis2003}. In this case the role of
the surface tension is played by the cell cytoplasmatic membrane.

The paper structure is as follows: In the next section the M\&M model is described and briefly derived so that the actions on the droplet of
its different terms can be understood. As stated before,  the shape dynamics depends upon the Lagrangian time history of the strain and rotation rates of the surrounding turbulent fluid. Full direct numerical simulations (DNS)  of turbulence can provide such information \citep{bif_eulerian_jfm2010} under the assumption of one-way coupling. The latter has been used extensively to study   particle relative dispersion \citep{Becetal-jfm-2010}, Lagrangian statistics properties of turbulence \citep{Meneveau11} 
and the fate of non-isotropic particles in turbulence \citep{Parsaetal12,ChevillardMeneveau13}.  In \S \ref{sec-DNSanalysis}
 we  simulate an ensemble of droplets, each drop obeying the M\&M model following the Lagrangian trajectories of fluid particles in DNS, at two Reynolds numbers. Statistical characterizations of resulting droplet sizes are provided based on the PDF of the largest diameter (the ellipsoid's largest principal axis). Particular attention is placed on the tails of the distributions,  to explore the fate of the most deformed droplets and how often these phenomena occur.  
Viscous drops deform because of hydrodynamical stresses and tend to maintain their shape owing to surface tension.  Polymers, described by purely elastic springs, share similar  characteristics and analogies with the case of polymers will thus prove useful. Polymers in turbulent flows in fact may undergo the so called coil--stretch transition if the local straining exceeds the restoring spring force of the polymer for a time long enough during the particle evolution. Such transition can be described by the tendency toward an unbounded growth of the polymer conformation tensor for continuum models as, e.g.,the Oldroyd-B \citep{balkovsky-etal-lungo,boffetta-etal}.   
 
 In \S \ref{sec-Cramer}   the PDFs of droplet sizes are related to 
the large deviation theory of the largest Finite Time Lyapunov Exponent for the case of unity viscous ratio. The formalism can be used to make quantitative predictions of the critical capillary number above which moments of the droplet size distribution diverge. The results from DNS are compared with these theoretical predictions. 

Besides the droplet size distribution, one is also interested in statistical characterizations of droplet shapes and orientations with respect to the local flow. Such properties are evaluated based on DNS and the M\&M model, and are presented in \S \ref{sec-DNSanalysis-geometry}. Variations of the ratio of droplet to carrier fluid viscosities are examined in \S \ref{sec-viscosityratio}.  Conclusions are presented  in \S \ref{sec-conclusions}. 

 \section{Lagrangian model for viscous, tri-axial ellipsoidal droplets in viscous shear flow} 
 \label{sec-Maffettonemodel}
 The model proposed by  \cite{MaffettoneMinale98} considers a drop of a viscous Newtonian fluid 
 immersed in another viscous Newtonian liquid of the same density,  subjected to flow such that an ellipsoidal drop shape of constant volume is maintained at all times. Of course, at significant deformations and especially close to the break-up the ellipsoidal shape is lost. However, some results \citep{GuidoVillone} support the idea that deformations away from ellipsoidal shapes develop only close to the critical shape. Under the assumption of ellipsoidal shape, the drop morphology and orientation can be entirely described by a positive--definite  second--order tensor $M_{ij}$. The tensor ${\bf M}$ is symmetric and its three eigenvalues correspond to the square of the semi--axes length while the eigenvectors give the orientations of the ellipsoid's axes. 
It can be understood as the inertia tensor of a droplet with constant density $\rho_d$: 
\begin{equation}
M_{ij}({\bf X}(t),t) = \rho_d \int_V  (r_i-X_i(t)) (r_j-X_j(t)) dV,
\label{eq:inertia}
\end{equation}
where the integral is extended over the whole volume of the droplet around the instantaneous position of  its center of mass, ${\bf X}(t)$.

In the M\&M model the drop deformation and  orientation dynamics are modeled using the rotation and strain rate of the underlying flow field (whose velocity components are $u_i$),  $\Omega_{kj} = 0.5(\partial_j u_k -\partial_k u_j)$ and  $S_{kj}=0.5(\partial_j u_k +\partial_k u_j)$ 
as:
 \be
 \frac{dM_{ij}}{dt} = \Omega_{ik}M_{kj}-M_{ik}\Omega_{kj}+f_2(\mu)(S_{ik}M_{kj}+M_{ik}S_{kj})-\frac{f_1(\mu)}{\tau}(M_{ij}-g(II_M,III_M) \delta_{ij}),
\label{eq:Maffettone}
 \ee
 where $f_1$ and $f_2$ are two functions that depend upon $\mu=\mu_i/\mu_o$ (the ratio of viscosities of the inner, $\mu_i$, and outer, $\mu_o$,
 fluids),  $\tau=\mu_o R/ \Lambda$ is the drop/bubble shape relaxation time-scale and $R$ the initial radius of the droplet (which is assumed spherical, initially).
In equation (\ref{eq:Maffettone}) the first two terms on the right hand side stem from the local rotation rate while the terms multiplied by $f_2$ define
the stretching due to viscous forces. The last term, proportional to $f_1$, models the tendency to restore the spherical shape induced by surface tension 
effects. 
Also, \be
g(II_M,III_M) = 3 ~\frac{III_M}{II_M}
\ee
enforces exact conservation of the droplet volume at all times as demonstrated by  \cite{MaffettoneMinale98}. 
The  factor $g(II_M,III_M)$ depends upon the 
invariants  of ${\bf M}$:
\be 
I_M = M_{kk}, ~~~~ II_M = -\frac{1}{2}\left( M_{ij}M_{ij} - I_M^2 \right), ~~~~ III_M = \frac{1}{3}\left( M_{ik}M_{kj} M_{ji} -  I_M^3 + 3 I_M II_M  \right).
\ee
Possible expressions for the rotation and stretching prefactors $f_1,f_2$ which match the known exact asymptotic limits
for small $Ca$, for infinite viscous ratio $1/\mu \to 0$ and for $\mu=1$ are given by \cite{MaffettoneMinale98} as:
\be
f_1(\mu) = \frac{40(\mu+1)}{(2\mu+3)(19\mu+16)}; \qquad f_2(\mu)= \frac{5}{2\mu+3}.
\ee
A number of other droplet models exist and \cite{Minaleetal10} provides a review 
of  the many other approaches  available to predict droplet dynamics and deformations in viscous flows.  
Here we use the above  M\&M model because of its relative simplicity and successful testing 
under various smooth flow conditions \citep{Guidoetal00,Minale04,Minale08,Minaleetal10}.
For neutrally buoyant small droplets placed in a turbulent flow, the Lagrangian evolution of equation (\ref{eq:Maffettone})
must be solved together with the droplet  position advected as a fluid particle. 
 
For future reference and convenience, we also provide a dimensionless version of the M\&M model that uses the velocity gradients (i.e. the small-scale turbulence inverse time-scale) to normalize time, and the initial droplet size to normalize length-scales (although since the dynamics are homogeneous with ${\bf M}$, its non-dimensionalization is not
relevant).  The reference inverse turbulent time-scale we use is defined as
\be
G_t= \left< \left( \frac{\partial u_1}{\partial x_1}\right)^2 \right>^{1/2}.
\ee 
Also, we define a capillary number according to 
 \be
\label{eq:ca}
 Ca = \frac{\mu_o R G_t}{\Lambda} = \tau ~ G_t  .
 \ee
Defining $S_{ij}^\prime = S_{ij} / G_t$,
$ \Omega_{ij}^\prime =  \Omega_{ij} / G_t$, 
 and $t^\prime = t G_t$, the equations are written in dimensionless form as follows:
 \be
 \frac{dM_{ij}}{dt^\prime} =  \left[ f_2(S^\prime_{ik}M_{kj}+M_{ik}S^\prime_{kj}) + \Omega^\prime_{ik}M_{kj}-M_{ik}\Omega^\prime_{kj}\right] ~ -~  \frac{f_1}{Ca}  \left(M_{ij}-3~ \frac{III_M}{II_M}  \delta_{ij} \right).
\label{eq:Maffettonenondim}
 \ee
Another characteristic time--scale exists, the Lagrangian correlation time of the velocity gradient tensor elements. This correlation
determines the temporal persistence of the applied straining and rotation rates. It is known that the correlation time scales of the strain--rate and vorticity differ
\citep{Gualaetal07,Yeungetal07,YuMeneveau10a,YuMeneveau10b} but both are known to scale with the Kolmogorov time--scale. Hence, they are of the order of $1/G_t$, but possibly with a large prefactor in the case of vorticity. 

 Among others,  we are interested in determining whether there is a steady state solution for the ``size'' of the droplets 
 defined in terms of the three semi--axes of its ellipsoidal shape. 
 Let us denote  the eigenvalues of ${\bf M}$ as $d_1^2$, $d_2^2$ and $d_3^2$, ordered according to $d_1>d_2>d_3$ and where $d_1$, $d_2$ and $d_3$ are the ellipsoid's semi--axes.  We recall that the  volume constraint implies that $d_1 d_2 d_3$ (strictly speaking the determinant ${\rm det}({\bf M}) = d_1^2 d_2^2 d_3^2$) remains constant in time.  
For large deformations, i.e. $d_1 >> d_3$, the trace of ${\bf M}$ ($I_M = d_1^2+d_2^2+d_3^2$) provides information essentially on the largest semi--axis.   
 
\section{Results from Direct Numerical Simulations}
\label{sec-DNSanalysis}
In this section, numerical solutions of equation (\ref{eq:Maffettone}) are presented.  As mentioned in the previous section,
we consider the case of droplets with a size much smaller than the viscous scale and with a negligible 
mismatch in density with the surrounding fluid. Under these conditions, 
the droplet center of mass evolves as a passive tracer in
the fluid and we can extract the time history of the velocity gradients along the Lagrangian 
trajectories of point--like particles following the equation:
\be
\dot {\bm X} \, =\, {\bm u}({\bm X}(t),t),
\label{eq:1}
\ee
where the Eulerian flow evolves according to the three dimensional Navier-Stokes equations:
\begin{equation}
  \partial_t\bm u + \bm u \cdot \nabla \bm u = -\nabla p +
  \nu\nabla^2\bm u +\bm F,\quad \nabla\cdot\bm u = 0\,.
  \label{eq:ns}
\end{equation}
The Lagrangian signals for the velocity gradient time histories, $\partial_j u_i({\bm X}(t),t)$, 
are obtained from DNS of homogeneous isotropic turbulence at two Reynolds numbers. 
The details about the DNS are given in table (I) 
(more details about the statistical properties of the Eulerian and Lagrangian fields can be found  in  \cite{Becetal-jfm-2010,Cencinietal-jot-2006}):
\begin{table*}
  \begin{tabular}{cccccccccccc} \hline\\[-13pt]
    & $N$ & $Re_{\lambda}$ & $\eta$ & $\delta x$ & $\varepsilon$ &
    $\nu$ & $\tau_{\eta}$ & $t_{\mathrm{dump}}$ & $\delta t$ & $T_L$ & $G_t$
    \\[-0pt]\hline Run I& 512 & 185 & 0.01 & 0.012 & 0.9 & 0.002 &
    0.047 & 0.004 & 0.0004 & 2.2 & 5.48 \\ Run II& 2048 & 400 & 0.0026 & 0.003 &
    0.88 & 0.00035 & 0.02 & 0.00115 & 0.000115 & 2.2 & 11.4 \\[-0pt]\hline
  \end{tabular}
  \caption{\label{table}
Eulerian parameters for the two set of data from the DNS of homogeneous and
isotropic turbulence. $N$ is the number of grid
    points in each spatial direction; $Re_{\lambda}$ is the
    Taylor-scale Reynolds number; $\eta$ is the Kolmogorov dissipative
    scale; $\delta x=\mathcal{L}/N$ is the grid spacing, with
    $\mathcal{L}=2\pi$ denoting the physical size of the numerical
    domain; $\tau_\eta =\left( \nu/\varepsilon \right)^{1/2}$ is the
    Kolmogorov dissipative time scale; $\varepsilon$ is the average
    rate of energy injection; $\nu=\mu_0/\rho$ is the kinematic viscosity;
    $t_{\mathrm{dump}}$ is the time interval between two successive
    data recordings along particle trajectories; $\delta t$ is the time 
    step of the model integration; 
    $T_L=L/U_0$ is the eddy turnover time at the integral scale $L=\pi$, 
    $U_0$ is the typical large-scale root-mean-square velocity and $G_t$ is the reference inverse turbulent time scale.
Averages are performed over 
two large eddy turnover times.
}
\end{table*}
The statistically homogeneous and isotropic external forcing $\bm F$
injects energy in the first low wave number shells, by keeping
constant their spectral content \citep{Chenetal1993}.
The kinematic viscosity $\nu=\mu_0/\rho$ is chosen such that the Kolmogorov
length scale $\eta\approx \delta x$, where $\delta x$ is the grid
spacing: This choice ensures a good resolution of the small--scale
velocity dynamics. The numerical domain is cubic and $2\pi$--periodic
in the three directions of space. We use a fully dealiased
pseudospectral algorithm with 2$^{\mathrm{nd}}$ order Adam--Bashforth
time--stepping (for details see \citep{Becetal-pof-2006,Cencinietal-jot-2006}). We analyze data from 
two series of DNS: Run I with numerical resolution of $512^3$ grid
points, and the Reynolds number at the Taylor microscale $Re_\lambda
\approx 185$; Run II with $2048^3$ resolution and $Re_\lambda \approx
400$. Particle trajectories are recorded at a frequency of 
$t_{dump} \sim \tau_\eta/10$ and followed for a total time of the order 
of  $2 T_L$, with $T_L$ the large eddy turnover time of the turbulent flow. 
We analyze a total of 
$15\times 10^3$ and $7 \times 10^3$ trajectories for Run II and Run I, respectively. The integration of
equation (\ref{eq:Maffettone}) is further refined by making a linear time interpolation  by a factor $10$ between 
two successive recorded data points along the Lagrangian trajectory. 

The evolution of the morphology tensor is stopped when the maximum deformation, defined as the
ratio between $d_1/d_3$ exceeds a factor $10^3$.  While this is
an arbitrary criterion to define a threshold 
associated with possible subsequent droplet breakup, it has been 
verified by various numerical tests (see below) that the dominant features of droplet dynamics 
and statistics show little sensitivity to the threshold value.
More advanced criteria taking into account droplet shape instabilities could be used, but especially under the limiting assumption that the 
shape remains ellipsoidal, this simple type of criterion is deemed appropriate for the focused objectives of this study. 

The initial drop size is assumed to be sufficiently smaller to the Kolmogorov scale $\eta$ so that even after severe deformations the largest scale still falls within the viscous range. The cut-off ratio $d_1/d_3=10^3$ means, due to the volume conservation, that $d_1$ is at most a factor 100 times the original scale. In practice the viscous range extends to scales of the order of $10\eta$ so that effectively, we are assuming that the initial scale is smaller or equal to $\eta/10$.
Note that due to the homogeneity of Eq. \ref{eq:Maffettone}, we can rescale $M_{ij}$ using the initial radius as characteristic length-scale without modifying the equation. Therefore the initial physical length-scale of the droplet does not explicitly enter into the dynamics, except through the relaxation time-scale $\tau$.

\begin{figure}
\includegraphics[width=14cm]{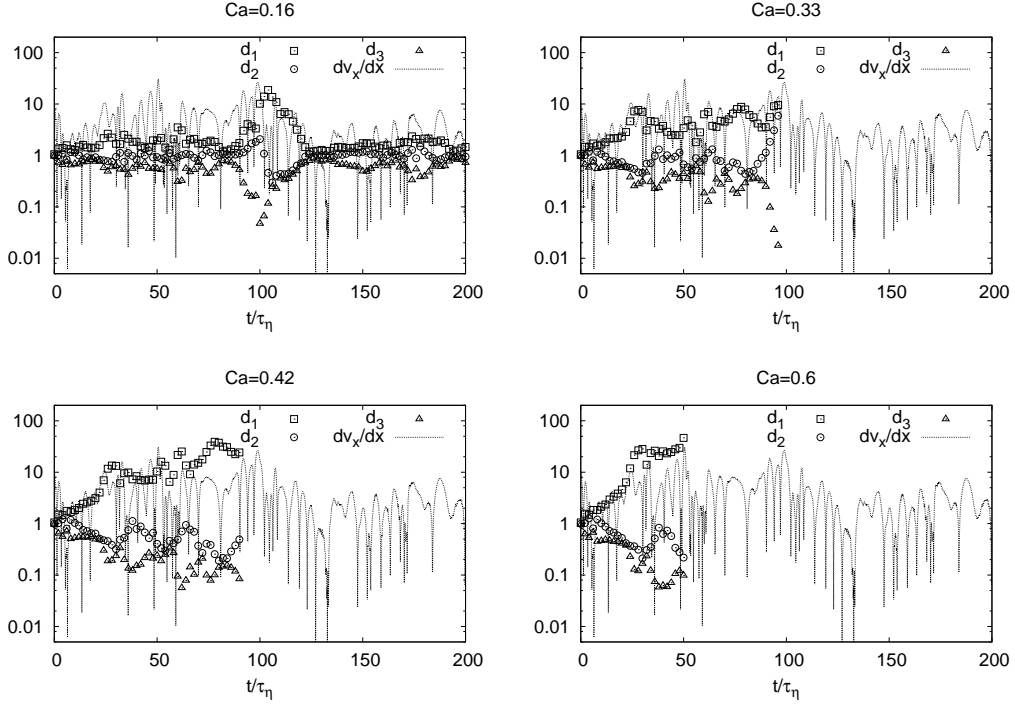}
\caption{\label{fig-signals-DNS}  Time signals of the square--root of the eigenvalues, 
$d_1(t) (={[\lambda_m^{(1)}]}^{1/2})$,  $d_2(t)$ and $d_3(t)$ in logarithmic units obtained from solving the 
Maffettone \& Minale droplet model coupled to Lagrangian time history of turbulent velocity gradients from DNS at $Re_\lambda =185$. 
We also superpose the time history of $|A_{11}(t)|$, the absolute value of one component of the 
velocity gradient tensor (solid line). 
}
\end{figure}

We begin by showing some typical time evolutions of droplets for different values of the  Capillary number.
In figure \ref{fig-signals-DNS} we show the time evolution of the square--root $d_i(t)=(\lambda_m^{(i)})^{1/2}$ of the three morphology tensor eigenvalues 
 for different relaxation times $\tau$, and for $\mu=1$, along a sample droplet trajectory. 
As one can see, at increasing relaxation time, i.e.  increasing 
Capillary number for a given turbulence intensity, the droplet tends to deform more and more. For instance,
the time history here represented shows a peak in the deformation at a time $t \sim 100 \tau_\eta$ during the droplet evolution 
where only the droplet with $Ca=0.16$ survives since it did not exceed the $d_1/d_3=10^3$ threshold.  

\begin{figure}
\includegraphics[width=14cm]{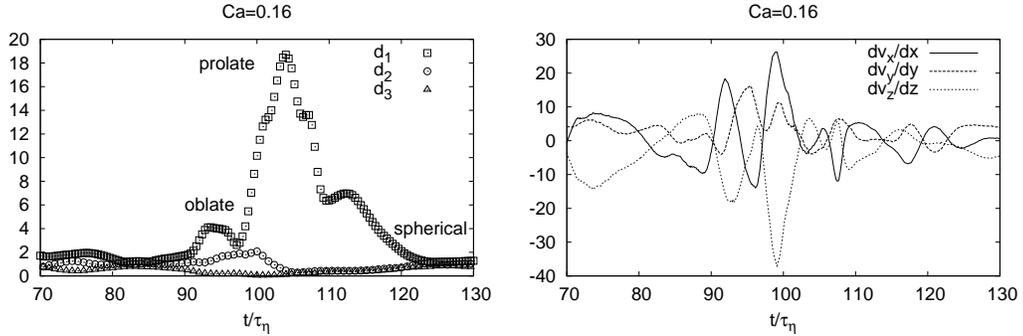}
\caption{\label{fig-signals-DNS2} Left: 
Enlargement of the previous panel for $Ca=0.16$ around the peak of the deformation of the 
droplet. 
Right: the corresponding evolution for the three diagonal entries of the velocity gradient tensor, 
$A_{11}(t), A_{22}(t), A_{33}(t)$.}
\end{figure}

In figure \ref{fig-signals-DNS2} we show an enlargement of this event, where 
one can better see the transition from an oblate to a strongly prolate shape during the droplet evolution. 
Only later the surviving droplet recovers a spherical--like shape with $d_1\sim d_2\sim d_3 \sim 1$.  In the same figure, on the right panel 
we show the time history of the three diagonal components 
of the velocity gradient tensor, in order to highlight the noticeable correlation between events where the droplet is strongly deformed and 
the underlying turbulent fluctuations of the fluid velocity field rate of deformation. As one can see, the event around $t \sim 100 \tau_\eta$,
where droplets with large Capillary number break, is preceded by strong intense oscillations in the turbulent stretching rates. 

\begin{figure}
\includegraphics[width=14cm]{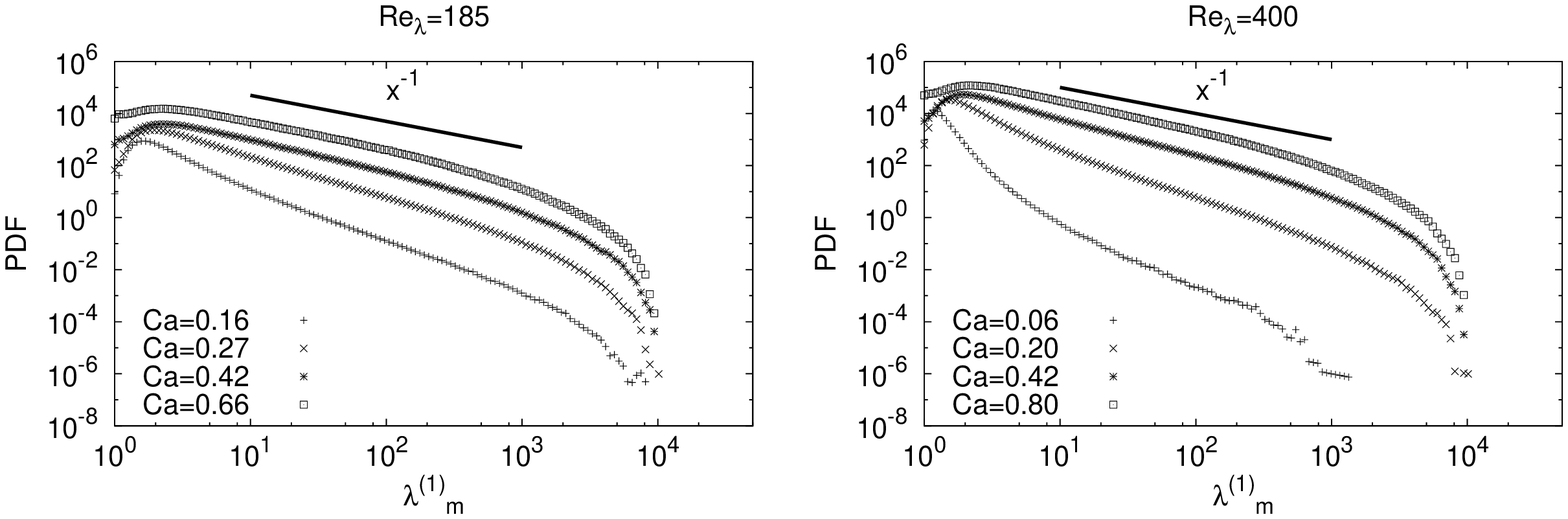}
\caption{\label{fig:1luca512}  Probability density functions of largest eigenvalue $\lambda_m^{(1)}$ of the morphology tensor ${\bf M}$, obtained from solving the Maffettone \& Minale droplet model coupled to Lagrangian time history of turbulent  velocity gradients from DNS at $Re_\lambda =185$ (left) and $Re_\lambda=400$ (right). 
Different curves correspond to different values of the capillary number,  both below and above the critical value, $Ca_c \sim 0.42$. 
The solid line represents the power law behaviour $\propto x^{-1}$.  Curves are shifted vertically for the sake of presentation.}
\end{figure}

As a next step, we focus on the distribution of droplet sizes. 
Figure \ref{fig:1luca512}  shows the PDFs of the largest ${\bf M}$ eigenvalue, $\lambda_m^{(1)}= d_1^2$, 
for various values of $Ca$ for both runs at the two Reynolds numbers.
As can be seen for increasing Capillary number, when the surface tension is decreased (for a fixed mean 
turbulent straining time-scale), longer tails develop and 
very large droplet deformation can occur. The tails approach a power--law form $P(x) \sim x^{-q}$ 
with $q$ decreasing (i.e. the tails becoming less steep)
at increasing capillary number. Similar behaviour can be observed at larger Reynolds numbers as shown in 
the right panel of the same figure.  

At capillary numbers above a threshold value, it is apparent that the PDF develops a $-1$ power--law tail, $q \to 1$. 
 
For such tails, while for finite size systems we may
operationally measure the PDFs due to the finite threshold imposed 
(we cannot exceed a maximum ratio between the highest and the lowest eigenvalue), if one were to imagine a system without
such cutoffs, the PDF could not be normalized because its integral diverges at large values. 
The transition to this behaviour appears to occur near $Ca_c \approx 0.4$ for both $Re_\lambda=185$ and $Re_\lambda =400$. 
We interpret such transition as follows: for $Ca>Ca_c$ if one waits long enough, with probability equal to unity all droplets would break eventually. For smaller $Ca$, breaking is still possible for some droplets (as, e.g. shown in figure \ref{fig-signals-DNS}), but large deformations become exponentially less probable.
In the next section an attempt will be made to predict this critical value of $Ca$ based on knowledge about 
turbulence small--scale statistics. 

\begin{figure}
\includegraphics[width=14cm]{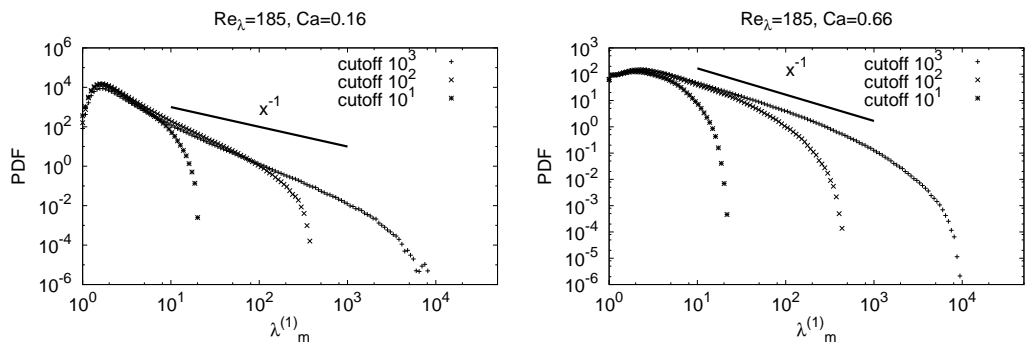}
\caption{\label{fig:check}  Probability density functions of largest eigenvalue $\lambda_m^{(1)}$ of the morphology tensor ${\bf M}$, obtained from solving the Maffettone \& Minale droplet model coupled to Lagrangian time history of turbulent  velocity gradients from DNS at $Re_\lambda =185$. Different symbols denote different  imposed cutoff values for the maximum deformation value $d_1/d_3 = 10, 10^2, 10^3$. Results are shown for two Capillary numbers $Ca= 0.16$ below (left) and above (right) $Ca_c$.
We also superpose  the power law behaviour $\propto x^{-1}$ predicted for the saturation slope when $Ca >Ca_c$.}
\end{figure}

In figure \ref{fig:check} we show the dependence of the PDFs on the $d_1/d_3$ threshold chosen as a criterion for droplet breakup, i.e. as a rule to stop the droplet trajectory. 
As is quite apparent, increasing the threshold leads to longer and longer tails of the PDF  without changing its main features. 
Using high threshold values  enables us to exhibit the power--law scaling of the PDF that 
develops for large scale disparities $d_1/d_3$.

\section{Cramer function formalism and finite--time Lyapunov exponents}
\label{sec-Cramer}
\begin{figure}
\centering
\includegraphics[width=10cm]{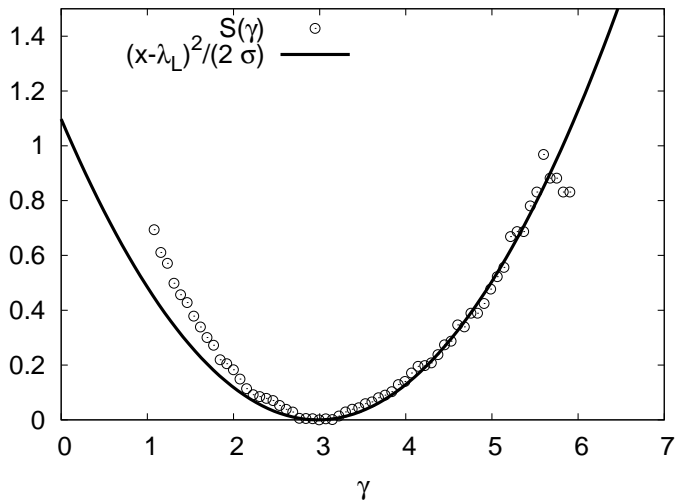}
\caption{\label{fig:cramer}  
Cramer function (from  \cite{Becetal-pof-2006}). We show also the  parabolic fit that uses 
$\lambda_L =0.14/\tau_\eta = 2.97$ 
and $\sigma = 0.19/\tau_\eta=4.04$  with  $\tau_\eta =0.047$. }
\end{figure}

 In this section we aim at establishing a semi analytical tool to 
predict the PDF shape of the maximal elongation of the droplet in the case when
the viscosity ratio is unity, $\mu=1$. In such a case, the stretching and the rotation terms of the M\&M model coincide with the ones describing the stretching and 
rotation of an infinitesimal volume of the fluid and therefore can be connected to the statistics of the
Lyapunov exponent of the particle trajectories inside the turbulent 
flow \citep{Becetal-pof-2006}. It is useful in this case to return to the 
dimensional description,  equation (\ref{eq:Maffettone}), in order to have a clearer understanding of the physical origin of all terms. When $f_2=1$  (viscosity ratio unity) the rotation rate 
and deformation rate tensors sum up to give the following evolution for the morphology tensor of the droplet:    
 \be
 \frac{dM_{ij}}{dt} = (A_{ik}M_{kj}+M_{ik}A_{kj}) ~ -~  \frac{f_1}{\tau} \left ( M_{ij}-g(III_M,II_M)  \delta_{ij} \right ) ,
\label{eq:Maffettone_visco1}
 \ee
where everything is expressed in terms of the velocity gradients, $A_{ik} = \partial u_i/\partial x_k$. 
It is useful here to point out that the evolution given by (\ref{eq:Maffettone_visco1}) 
is very close to the one of polymer stretching by a turbulent flow as for the case of the approximation given by 
the  Oldroyd-B model.
In the latter the role played by the morphology tensor ${\bf M}$ is played by the  polymer 
conformation tensor $C_{ij}({\bf X}(t),t) = \overline{R_i R_j}$  
(where ${\bf R}$ is  the ``ends---to---ends polymer vector'' and the average is intended over the thermal noise applied 
 to each molecule inside an infinitesimal volume advected by the flow, see 
\cite{balkovsky-etal-lungo,Chertkov2000} 
for a rigorous discussion). For the polymer case, the linear damping is given by a relaxation time--scale toward the equilibrium isotropic extension.  The only difference between the two cases is that for polymer there is no need to enforce the  volume conservation and therefore the term $g$ is typically set to unity in an equation like Eq. 
\ref{eq:Maffettone_visco1}. For the tails of the probability distribution of the largest eigenvalue of ${\bf M}$, these differences are expected to have negligible effects since
along the tail we have $Tr({\bf M}) >>1$ and in that limit $M_{ij}-g\delta_{ij} \approx M_{ij}$ ($g$ tends to zero since $II_M$ grows while $III_M$ remains unity). 
The long time evolution of the droplet morphology tensor, given by 
 (\ref{eq:Maffettone_visco1}), 
will depend critically upon
the balance between two different mechanisms: The first is given by the accumulation of the stretching effects induced by the underlying flow, as expressed by the terms $(A_{ik}M_{kj}+M_{ik}A_{kj})$  on the right hand side of equation (\ref{eq:Maffettone_visco1}). The second mechanism is given by the relaxation toward an isotropic configuration as expressed by the term $(f_1/\tau)  (M_{ij} -g \delta_{ij})$. If the former is strong enough to  dominate the long term behaviour, the droplet will be in a {\it stretched} configuration (and it keeps stretching with one or two of its length-scales growing in an unbounded fashion if not resisted by additional nonlinear stiffness mechanisms.)
In the opposite case it will be, on average, in a {\it coiled} configuration, where we have used this term to stress the analogy with the polymer dynamics. In order to predict the critical capillary number where  stretching will overwhelm the surface tension effects it is possible to apply the same balance 
 already successfully used for the polymer case by \cite{balkovsky-etal-lungo,boffetta-etal}. 
The idea is to control the  asymptotic behaviour of the trace of the
 morphology tensor, $ Tr({\bf M}(t))$ observing that thanks to (\ref{eq:inertia}) 
 it is equivalent to the tensorial product of two infinitesimal  vectors
defining the position of a generic particle inside the droplet, $R_i = x_i-X_i(t)$. We therefore can restrict ourself to study the evolution of a fluid line element in the fluid:
\be
\label{eq:lyapunov0}
 \frac{dR_{i}}{dt} = A_{ik}R_{k}
\ee
and then taking the square of it in a suitable sense.  To characterize  the long-time evolution of (\ref{eq:lyapunov0}) it is useful 
to introduce the so-called finite--time Lyapunov exponents (FTLE):
\be
\gamma(t) = \frac{1}{t} log \left ( \frac{|{\bf R}(t)|}{|{\bf R}(0)|} \right ) ,
\label{eq:lyapunov}
\ee
which for large $t$ tends   with probability one to the 
largest Lyapunov exponent governing the chaotic properties of particles trajectories in the turbulent flow, 
$\lambda_L = \lim_{t\rightarrow \infty} \gamma(t)$.
 However, if we do not perform the limit $t \rightarrow \infty$, the FTLE  
exhibits deviations from the mean. These fluctuations are described  by the 
probability distribution
 function of $\gamma$ at various times {\it via} the large deviation theorem \citep{Frisch95}: 
\be
P(\gamma,t)  \sim   \exp( - t S(\gamma)),\qquad t \to \infty .
\label{eq:cramer}
\ee
The function $S(\gamma)$ is the so--called Cramer function (see \cite{Frisch95} for a text book introduction and 
also \cite{Paladin87,Eckmann86}) which must be convex, semi--positive definite and vanishing  at $\gamma= \lambda_L$, because for $t \rightarrow \infty$ we must have with probability one that the FTLE converge
to the largest  Lyapunov exponent,
 $\lambda_L$. 
Combining  equations (\ref{eq:lyapunov}) and (\ref{eq:cramer})  one may write for the $q$th-order moments of the vector growth:
\be
\label{eq:FTLE}
\left< \left(\frac{|{\bf R}(t)|}{|{\bf R}(0)|} \right)^q \right> = \int  ~ \exp[t\, (q \gamma -S(\gamma))] ~ d\gamma ~\sim ~ \exp(t\, L(q)),
\ee
where the last passage is obtained in the saddle point  limit of large $t$ subject to the condition: 
\be
L(q) = \max (q \gamma - S(\gamma)).
\ee
Moreover, one can show that  the Lyapunov exponent $\lambda_L = L'(0)= \frac{d L(q)}{dq}|_{q=0}$.
We must now just notice that the stretching part of the evolution (\ref{eq:Maffettone_visco1}) is twice the right hand side of  
equation  (\ref{eq:lyapunov0}), i.e. $ M_{ij}  \sim 
R_iR_j$. As a result, we have that the large deviation properties of the largest
 eigenvalue of the morphology tensor are controlled by $L(2q)$ instead of $L(q)$.
Considering also the linear relaxation induced by the surface tension terms and neglecting the $O(1)$ terms in fronts to the $\delta_{ij}$, which must not be important when  $Tr({\bf M}) \gg 1 $, we end up with the prediction that  for large times:
\be
\label{eq:qcritical1}
\langle [Tr({\bf M}(t))]^q \rangle \sim 
\exp \left [ t \left (L(2q) -q \frac{f_1}{\tau} \right ) \right ] .
\ee
It is possible now to derive a criterion for the existence of a stationary probability distribution for the 
morphology tensor. A stationary  PDF  must be  normalizable at all times, i.e. the exponent
 $L(2q) -qf_1/\tau$ must be zero when $q =0$, such that $ \lim_{q \to 0} \int  x^q P(x) dx= 
\int P(x) dx =1$. The latter condition implies that there exists a critical relaxation time
$\tau^c$ such that:
\be 
\lim_{q \to 0}[L(2q) - q f_1/\tau^c] = 0, ~~~~~{\rm or} ~~~\tau^c = \lim_{q \to 0} ( f_1/L(2q)) = f_1/(2L'(0)) = f_1/(2\lambda_L). 
\ee  
For $\tau>\tau^c$ the  tensor does not reach a stationary distribution and it is indefinitely stretched. 
In that limit  we will have that all moments diverge (which corresponds to the Weissenberg criterion for the coil--stretched transition in the case of polymers).
For $\tau < \tau^c$, when the PDF of the trace,  $x = Tr({\bf M})$,  is normalizable, 
the tail  will scale like
\be 
P(x) \sim x^{-(1+\tilde q(\tau))}.
\ee
The critical exponent of the tail is given by the largest order of the moment that  does not diverge, i.e. $\tilde q(\tau)$ is such that 
\be 
\label{eq:qcritical2}
L(2\tilde q(\tau)) = \frac{f_1 \tilde q(\tau)}{\tau}.
\ee
For many practical purposes, the Cramer function $S(\gamma)$ can be expanded in Taylor series around its minimum up to second order,
\be
\label{eq:parabola}
S(\gamma) = (\gamma - \lambda_L)^2/(2\sigma),
\ee
where $\sigma$ is a parameter characterizing the degree of intermittency and variability of the FTLE. 
The Cramer function of Navier--Stokes turbulence in three dimensions has been
measured in prior work, based on Lagrangian tracking and integration of 
fluid velocity gradients from DNS at $512^3$ and $Re_\lambda = 185$ \citep{Becetal-pof-2006}.
In figure \ref{fig:cramer}  this measured Cramer function  is shown and the measured (fitted)  $\lambda_L$ and 
$\sigma$ are given in the caption. 
Let us first notice that the Cramer function cannot be exactly parabolic for all values of $\gamma$; this is due to the fact that the incompressibility constraint forces $\gamma >0$. Nevertheless, it is possible to find a good parabolic   fit for the right branch of the parabola, the only one that will be of interest for 
us because  the condition (\ref{eq:qcritical2}) gives values of  $\tilde q(\tau)$ that correspond to  $\gamma(\tilde q) > \lambda_L$  (see below). 

\subsection{Comparison with DNS}
In the DNS used to obtain the results shown in figure  \ref{fig:1luca512},   the parameters  $f_1 =0.457$ 
and $f_2=1$ (unity viscous ratio) were used. Also, for the DNS flow, the main Lyapunov exponent is $\lambda_L \sim 2.97$  \citep{Becetal-pof-2006}. 
Therefore we predict as a critical relaxation time scale the value
\be
\tau^c = f_1/(2 \lambda_L) \sim 0.077,
\ee
which in terms of the critical Capillary number means $Ca_c =
 \tau^c \langle (\partial u_1/\partial x_1)^2\rangle^{0.5} = 0.42$. 
This value of critical Capillary number or $\tau^c$ is very well confirmed by the results shown in figure \ref{fig:1luca512},   
where the PDFs were shown to  saturate to  a $-1$ tail  for $\tau \sim 0.077$ ($Ca_c = 0.42$) 
for the $Re_\lambda=185$ DNS case.
Because the dependency on the Reynolds number enters only via the Lyapunov exponent $\lambda_L$, 
which is known to follow  the relation $\lambda_L \sim 0.14/\tau_\eta$ \citep{Becetal-pof-2006}, it is also possible to 
predict the critical relaxation time for  the $Re_\lambda=400$  case. In particular, replacing the values for $\tau_\eta$ given in table (I)  
we must have $\tau^c \sim 0.033$ for $Re_\lambda=400$, which would correspond  to $Ca_c \sim 0.37$  again in 
good agreement with the observed accumulation of the PDF's tail 
shown in the  left panel of the same figure. Notice that the critical Capillary number should at first sight not depend on Reynolds
 number because the two quantities  $\lambda_L$ and  $\langle (\partial u_1/\partial x_1)^2\rangle^{0.5}$  have
the same Reynolds dependency on dimensional grounds. The numerical results shown in \cite{Becetal-pof-2006} suggest the presence of a small intermittency correction 
to the rule $\lambda_L \propto 1/\tau_\eta$.  Moreover,  in order to understand the Reynolds dependency of eq. (\ref{eq:ca})  one would need also to consider the intermittent corrections to the statistics of velocity gradients \citep{Benzi91}. It is difficult to say if the observed small dependency of $Ca_c$ on Reynolds is due to these  two combined intermittent corrections or it is just induced by small statistical fluctuations on the measured quantities. Data from a larger variation in Reynolds are needed in order to answer this important question. 

Next, we explore whether  relation (\ref{eq:qcritical2}) can be used to estimate the  PDF slopes for $\tau<\tau^c$, i.e. before
criticality.   Again, we use the published Cramer  function \citep{Becetal-pof-2006} (see figure \ref{fig:cramer}). Using a parabolic fit (\ref{eq:parabola}), 
the Legendre transform can be worked out analytically and the maximum is reached  for
\be
\label{eq:gammat}
\tilde \gamma(q) = \lambda_L + 2 \sigma q .
\ee
Moreover,
\be
L(2q) = 2 q \lambda_L + 2 \sigma q^2,
\ee
which gives the prediction for the slope (from \ref{eq:qcritical2}):
\be
\label{eq:ld}
\tilde q(\tau) =  \frac{f_1/\tau - 2\lambda_L}{2 \sigma}.
\ee
The first thing to be noticed is that from equation  (\ref{eq:gammat}), already for
 $q=0.5$ we have $\tilde \gamma(1) = \lambda_L + \sigma \sim 7$, 
i.e. already for $\tilde q \sim 0.5$ we are probing the far tails of the Cramer function (\ref{fig:cramer}). 
Hence, to remain within good statistical confidence, we can apply  the prediction (\ref{eq:ld}) only if it is satisfied for $\tilde q \ll 1$, i.e. 
for Capillary numbers and relaxation times relatively close to the critical ones.  
In tables (II) and (III) we report for both RUN I and RUN II the values for the slopes of the PDF tails  
using the expressions $\sigma = 0.19/\tau_\eta$ and $\lambda_L = 0.14/\tau_\eta$ which we showed in 
figure \ref{fig:cramer} to be a good fit for the right branch of the Cramer function. 
 \begin{table*}
 \begin{center}
 \begin{tabular}{| l |c| c| c| c|}
 \hline
                             & $\tau=0.03$ & $\tau=0.05$ & $\tau=0.06$& $\tau=0.077$  \\
                             & $Ca=0.16$ & $Ca=0.27$ & $Ca=0.33$& $Ca=0.42$  \\
 \hline
  $\tilde q(\tau)  $       &$1.15$        & $0.39$       & $0.21$      &     $0$ \\
 \hline
 \end{tabular}
\caption{Values of the asymptotic power-law slopes for the PDF of the trace of the morphology tensor
 for $Re_\lambda=185$ simulation, for these parameters we have $\tau^c \sim 0.077$ ($Ca_c \sim 0.42$)}
 \end{center}
 \label{table1}
\end{table*}

\begin{figure}
\centering
\includegraphics[width=10cm]{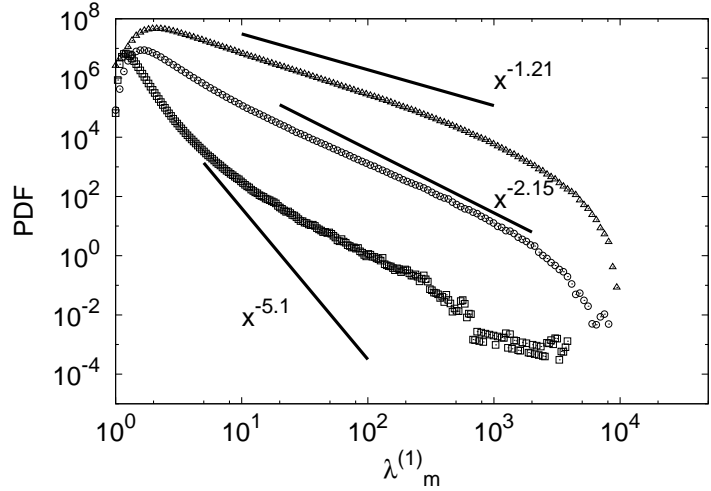}
\caption{ \label{fig:003} PDFs of the largest eigenvalue of the morphology tensor for $Ca=0.06$ (squares), $Ca=0.16$ (circles) and $Ca=0.33$ (triangles) as computed from the simulation. The  corresponding theoretical power law  predictions are shown as solid lines with the predicted slopes (exponents) noted.}
\end{figure}

 \begin{table*}
 \begin{center}
 \begin{tabular}{| l |c| c| c| c|}
 \hline
                                        & $\tau=0.005$ & $\tau=0.015$ & $\tau=0.025$& $\tau=0.033$  \\
                                        & $Ca=0.06$ & $Ca=0.2$ & $Ca=0.32$& $Ca=0.37$  \\
 \hline
  $\tilde q(\tau)  $       &$4.1$        & $0.87$       & $0.23$      &     $0$ \\
 \hline
 \end{tabular}
\caption{Asymptotic power-law slopes for the PDF of the trace of the morphology tensor for $Re_\lambda=400$ simulation. For these parameters we have $\tau^c \sim 0.033$,  ($Ca_c\sim 0.37$)}
 \end{center}
 \label{table2}
\end{table*}
In figure \ref{fig:003} we superpose the PDFs of the largest eigenvalue of the morphology
tensor ${\bm M}$  at different Capillary numbers for some
typical values given in tables (II) and (III). One can notice that by increasing the Capillary number,
the theoretical prediction based on the large deviation theory for the FTLE becomes increasingly better. 
The deviations for $Ca \ll Ca_c$ could be due to the following reasons: First, as said, for small $Ca$
the saddle point estimate is dominated by very large values of the FTLE, leading to a bigger 
statistical uncertainty. Second, if the Capillary number is small, the stretching terms are less important, the
morphology tensor is closer to its isotropic shape, stretching does not persist for long times and probably
the asymptotic estimate of the large deviation Cramer function is  not suitable for such intermediate
situations.
\section{Further characterization of droplet orientations and morphology}
\label{sec-DNSanalysis-geometry}

In this section, we provide results from the analysis of droplets in DNS not only focusing on the largest eigenvalue as done in the preceding sections, but also 
characterizing the droplet morphology and orientation dynamics.  In order to characterize the shape of the particle, we use the deformation parameter also used in \cite{Guidoetal00}:
\be 
P(D),~~~~~~{\rm where} ~~~~~D = \frac{d_1-d_3}{d_1+d_3}.
\ee
This parameter does not distinguish between disk--like and cigar--like shapes, but for a sphere one has $D=0$ and for the most deformed possible states (either disk-- or spaghetti--like), one has $D=1$. In addition, to distinguish also between disk--like and cigar--like shapes, the $s^*$ parameter introduced by \cite{LundRogers94} to characterize the rate of strain eigenvalues (that add to zero) can be used if properly modified. For this purpose,  the eigenvalues must first be re--defined in terms of logarithmic variables. We define
\be
r_i = \ln(d_i/d_i(0)),
\ee
then $\sum_{i=1}^3 r_i = 0$. We then define the  \cite{LundRogers94} parameter 
\be
s^*~=~-3 \sqrt{6} ~\frac{r_1 r_2 r_3}{(r_1^2+r_2^2+r_3^2)^{3/2}}
\ee
which is such that $s^*=+1$ indicates disk--like shapes while $s^*=-1$ indicates long fiber--like shapes. However, $s^*=0$ is somewhat indeterminate: 
it can mean either a sphere or an ellipsoid in which the intermediate axes remains undeformed
with $d_2(t)= d_2(0)$ leading to $r_2=0$ and $s^*=0$. Still, peaks of the PDF of $s^*$, $P(s^*)$, near either $s^*=\pm 1$ can be interpreted quite clearly.  

\begin{figure}
\centering
\includegraphics[width=10cm]{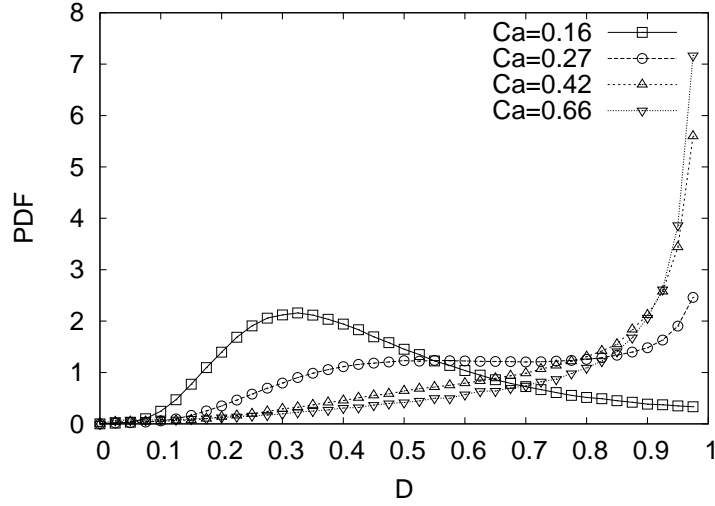}
\caption{\label{fig-PDF-MDee-Ca-all} PDFs of $D=(d_1-d_3)/(d_1+d_3)$, for different $Ca$.}
\end{figure}

\begin{figure}
\centering
\includegraphics[width=10cm]{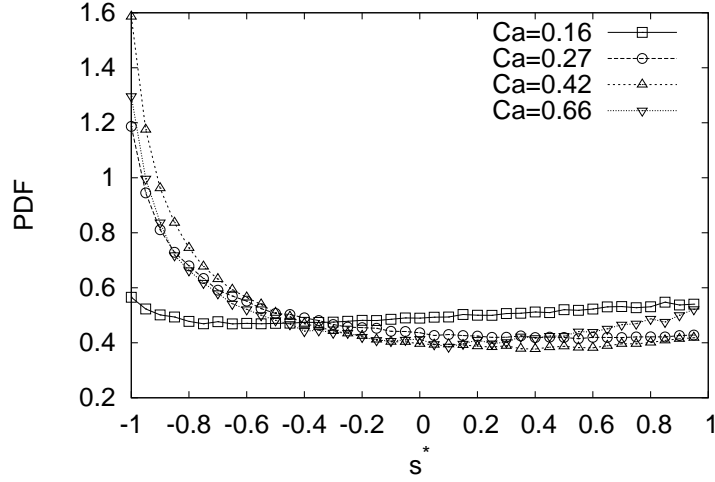}
\caption{\label{fig-PDF-Msstar-Ca-all} PDFs of $s^*$, for different $Ca$.}
\end{figure}

We present PDFs of the parameters $D$ and $s^*$ in figures \ref{fig-PDF-MDee-Ca-all} and  \ref{fig-PDF-Msstar-Ca-all}, respectively,
for various of the $Ca$ considered from the $Re_\lambda=185$ simulation. From the results for $D$ the trends are clear: for increasing 
capillary number, the droplets become more and more anisotropic, with an increasing ratio between largest and smallest principal axes. 
The trends shown in figure \ref{fig-PDF-Msstar-Ca-all} for $s^*$ are less monotonic.
There is a tendency toward rod--like shape for capillary numbers approaching the 
critical value, and a small recovery towards more disk--like shapes for very large Capillary numbers. 

Prior work has studied orientation dynamics and tumbling rates of non--deforming rigid ellipsoidal particles. 
In particular, e.g. works of  \cite{ShinKoch05,Parsaetal12,ChevillardMeneveau13}
show that  particles with one elongated direction and two small ones (fiber or rods---like, $s^*\to -1$) tend to align with the vorticity, which implies rotation around the major axis thus effectively reducing the tumbling rate of that major axis. As the anisotropy of the particle is increased, its tumbling rate is reduced. Conversely, \cite{Parsaetal12}  have found that the tumbling rates of disk--like particles (two large and one very small major axis, $s^* \to 1$) is significantly increased. This trend is due to the fact that  disk--like particles tend to align with the most contracting eigendirection of the strain--rate tensor which, in turbulence, happens to be preferentially orthogonal to the vorticity vector \citep{ChevillardMeneveau13}. The vorticity then spins the disk strongly, not 
unlike setting a coin spinning on a tabletop. However, if the particle is allowed to deform, these
trends can be expected to be modified significantly. Hence, the alignment properties of droplets with the strain--rate and vorticity are of considerable interest. 

\begin{figure}
\centering
\includegraphics[width=14.5cm]{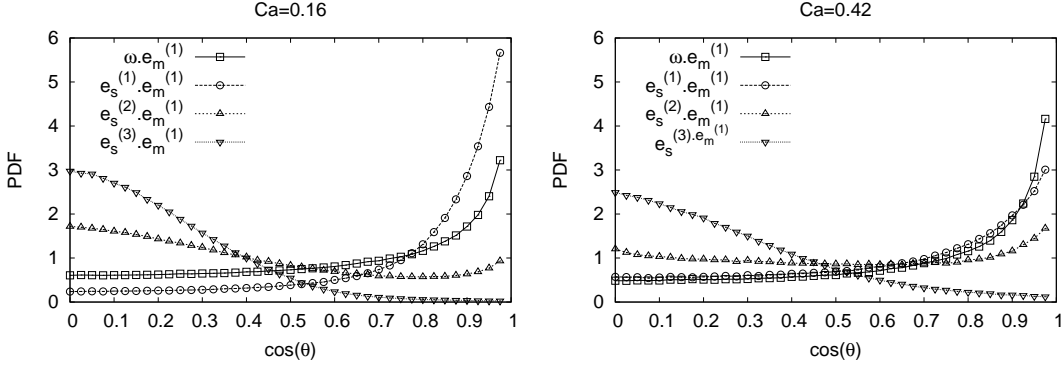}
\caption{\label{fig-PDF-Me1-Ca-all} PDFs of cosine of angle between the particle's largest eigenvector, ${\bf e}_m^{(1)}$, with vorticity and the 
three eigenvectors of ${\bf S}$, for two $Ca$ (left and right). }
\end{figure}

In order to characterize the orientation statistics of droplets relative to the flow field, we are interested in the PDFs of cosine of angles between, ${\bf e}_m^{(1)}$,  the eigenvector corresponding to the largest semi-axis of the morphology tensor and a few characteristic directions of the underlying flows
(we will consider only absolute values of the cosines to avoid problems with direction ambiguities):
\be 
P(| {\bf e}_m^{(1)} \cdot \hat{\vct{\omega}} |), ~~~~P(| {\bf e}_m^{(1)} \cdot {\bf e}_s^{(1)} |), ~~~~P(| {\bf e}_m^{(1)} \cdot {\bf e}_s^{(2)} |), ~~~~~P(| {\bf e}_m^{(1)} \cdot {\bf e}_s^{(3)} |), 
\ee
where $ \hat{\vct{\omega}}$ is the unit vector in the vorticity direction, and $ {\bf e}_s^{(k)}$ ($k=1,2,3$) are the three orthogonal strain--rate eigendirection unit vectors. In terms of the alignment of the ellipsoids relative to features of the turbulent flow, in  figures \ref{fig-PDF-Me1-Ca-all},
we show PDFs of the cosine of the angle with each of the four directions characterizing the local turbulent flow: 
the three strain--rate eigendirections and the vorticity for two different Capillary numbers.
From these results it is apparent that the droplet largest eigendirection tends to align with the most extensive strain--rate eigendirection, as well as with the 
vorticity. It also often aligns with the second intermediate strain--rate eigendirection (which itself is well aligned with the vorticity). Conversely, it tends to be orthogonal to the most contracting eigendirection.  At increasing Capillary number, the alignment is less pronounced, since the droplet has less time to 
synchronize with the underlying flow topology before it deforms and reaches the threshold deformation levels leading to eventual breakup. 
\section{Effects of the viscosity ratio}
\label{sec-viscosityratio}
\begin{figure}
\centering
\includegraphics[width=8cm]{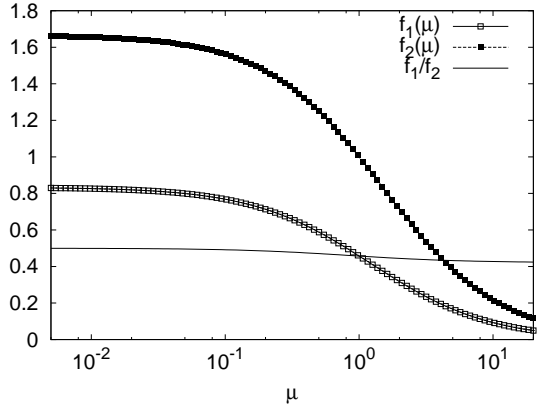}
\caption{
\label{fig:f1f2} Behavior of $f_1(\mu)$, $f_2(\mu)$ and of their ratio at changing $\mu$. 
Notice that for $\mu \ll 1$ the prefactor in front of the symmetric stress tensor  
term (the only one which stretches) becomes very small. As a result the deformation of the droplet is much smaller if
 the Capillary number is kept unchanged.}
\end{figure}

When changing the viscosity ratio $\mu = \mu_i/\mu_o$, we change the relative importance of stretching with respect to rotation 
in equation  (\ref{eq:Maffettone}).  In figure \ref{fig:f1f2} we show their functional dependency as a function of $\mu$, as given by the phenomenological  dependency of $f_2$ and $f_1$ proposed by \cite{MaffettoneMinale98}. We can see that while the ratio $f_1/f_2$ is always close 
to $0.45$, for large viscosity ratios the values of $f_2$ and $f_1$ can change up to a factor $4-5$. 

\begin{figure}
\centering
\includegraphics[width=14cm]{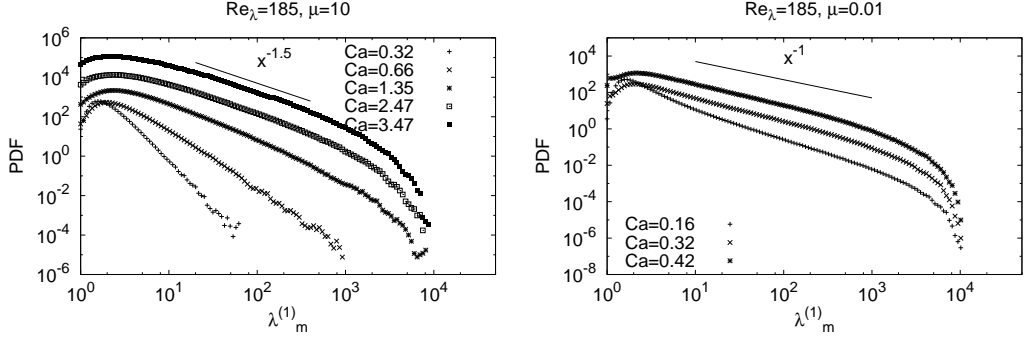}
\caption{Left panel: PDF of the largest eigenvalue of the morphology tensor for various Capillary numbers, for viscosity ratio $\mu=10$. 
Note that the straight line has a slope $-1.5$ and that the saturation of the tails happens at a higher $Ca$ value compared to
the critical Capillary estimated by (\ref{eq:mu10}). Right panel: the same for $\mu=0.01$. Now the transition is very similar to the case $\mu=1$, and the tail has the -1 power law slope.
\label{fig:fig0.01_10} 
}
\end{figure}

Next, it is of interest to attempt to apply our earlier Cramer--function predictions to the case of
different viscosity ratios. For instance,  for the case at $\mu=10$, loosely applying an `order--of--magnitude' estimate one would have predicted that 
the transition to a `non--stationary' regime (unbounded growth of the major axis or axes) would happen when:
\be
\label{eq:mu10}
f_2(10) \lambda_L  \sim f_1(10)/\tau_c.
\ee
This is because now the stretching part is proportional to $f_2$ and we are supposing that the symmetric stress tensor, $S_{ij}= (A_{ij}+A_{ji})/2$,
leads to the same Lyapunov exponent $\lambda_L$ as that of the original one, $A_{ij}$. If this were true, we should expect for 
the transition to occur at
 $Ca_c \sim 0.81$ (because from Eq. \ref{eq:mu10} we have $\tau^c = f_1(10)/f_2(10) \lambda_L^{-1} = 
 0.148$, and with $G_t=5.48$ we obtain $Ca_c=\tau^c G_t = 0.81$).
This, however,  is not what is observed in the numerical results shown in  figure \ref{fig:fig0.01_10}.
As one can see on the left panel of figure \ref{fig:fig0.01_10}, the saturation seems to be
present, but now at around $Ca_c \sim 2.5$, i.e. it is delayed. This might be understood heuristically 
by noting that the rotation is decorrelating the droplet orientation from the stretching rate thus making strong deformations less likely.
Moreover the slope of the PDF is not close to $-1$, meaning that the physics of the deformations and 
relaxations, and its relationships to the flow, differ significantly from the $\mu=1$ case. 

Similarly, on the right panel of the same figure the results for $\mu=0.01$ are shown. At small $\mu$ values, one
expects the effects of rotation to be negligible compared to the stretching and relaxation. As can be seen,  the transition happens
almost at a similar value of the case $\mu=1$ and the PDF has a characteristic $-1$ slope in this case. 

\begin{figure}
\centering
\includegraphics[width=10cm]{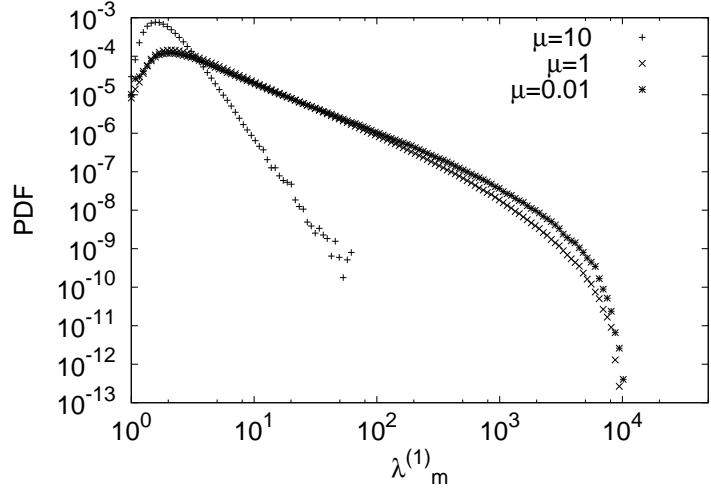}
\caption{PDF of the largest eigenvalue of the morphology tensor for  $Ca=0.32$ of the three viscous ratio, $\mu=0.01,1,10$}
\label{fig:3mu} 
\end{figure}

\begin{figure}
\centering
\includegraphics[width=14cm]{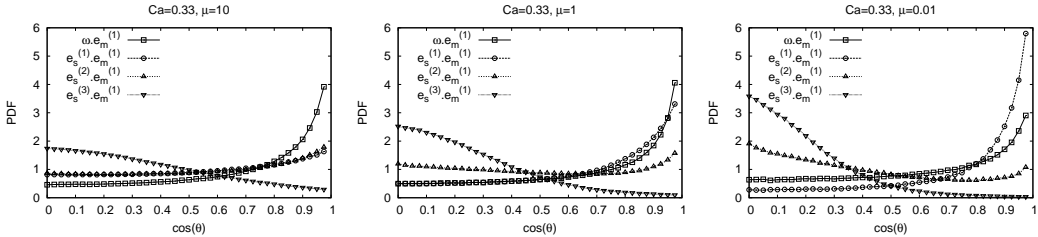}
\caption{PDFs of cosine of angles. Comparison between alignment with vorticity and strain-rate eigendirections for three different viscosity ratios,  at  $Ca=0.33$.}
\label{fig:omega} 
\end{figure}

\begin{figure}
\centering
\includegraphics[width=14cm]{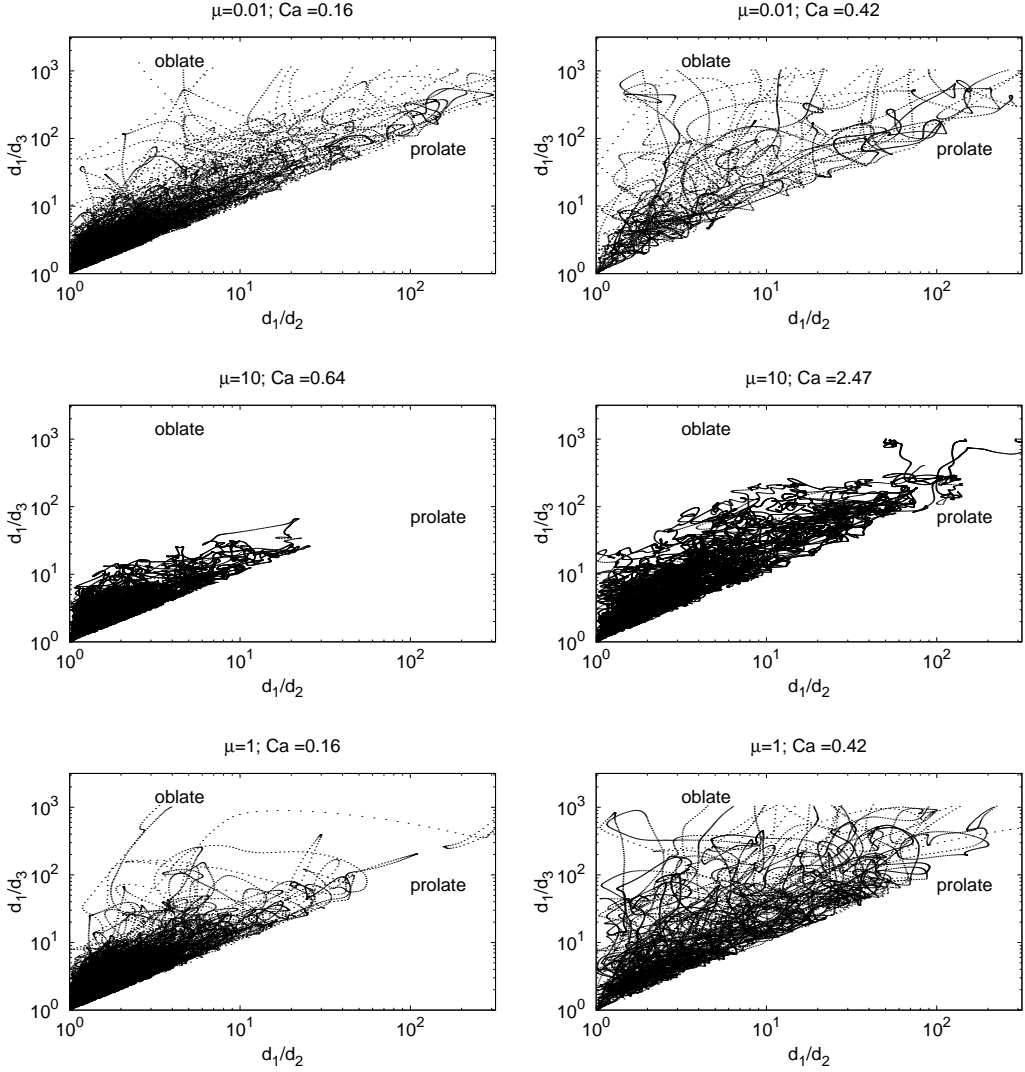}
\caption{Scatter plot of ratio between ellipsoidal semi-axes sizes.}
\label{fig:scatter} 
\end{figure}

In order to better highlight the dependency upon $\mu$, in figure \ref{fig:3mu} the
results are superposed for a fixed $Ca$  value for  three viscosity ratios
$\mu=0.01,1,10$. As one can see, the case at $\mu=0.01$ stretches slightly better 
than $\mu=1$ ($f_2$ is larger for $\mu=0.01$). On the other hand, as we knew already, the $\mu=10$ is very contracted, rotation here dominates.  
In figure \ref{fig:omega} the alignment between the maximum elongation and the vorticity and 
the strongest stretching rate is shown for different viscous ratios at $Ca=0.33$.
Notice that only for $\mu=1$ we have a very similar weight of rotation and stretching. For $\mu=10$ stretching is fully 
uncorrelated with the highest deformation, as expected. 
On the other hand for $\mu=0.01$  the elongation is more oriented with the strain rate.
In figure \ref{fig:scatter} we show  scatter plots of $d_1/d_2$ vs $d_2/d_3$ 
at changing $\tau$ and the viscous ratio $\mu$. Notice how the region corresponding to disks $d_1/d_2 \sim O(1)$ and
$d_1/d_3 \gg 1$ is strongly depleted for the case when $\mu =10$, i.e. when the stretching rate is not efficient. Evidently, in this case the 
droplets tend to be aligned with vorticity and become rod--like.
\section{Conclusions}
\label{sec-conclusions}
The statistical distribution of semi--axes scales and orientations of small ellipsoidal droplets (with a size smaller than the Kolmogorov scale) 
in fully developed homogeneous and isotropic turbulent flows has been studied. 
Droplets are supposed to be fully passive and diluted (no droplet--droplet interactions). In the limit of very small size, 
droplets can be considered inertialess with their centre of mass following the trajectories of a fluid tracer. 
Deformation induced by turbulent strain-rate and rotation is studied by means of a simplified model proposed by \cite{MaffettoneMinale98} that considers droplets to remain of ellipsoidal shape and including a restoration force due to surface tension effects that conserves droplet volume.  
A critical capillary number is identified at which one obtains unbounded droplet growth along one or two directions (which eventually should lead to droplet breakup). 
At unity droplet--to--fluid viscous ratio,  one can exploit analogies with polymers to obtain analytical predictions of 
the critical Capillary number as a function of the largest Lyapunov exponent of the trajectories of fluid particles and the relaxation time--scale.
Large deviation theory for the largest Finite Time Lyapunov exponent allows to predict also the power law tail of the PDF of the largest droplet dimension. 
Another interesting question is connected with the temporal properties of the droplets dynamics. In order to determine a break-up frequency  one needs to study numerically the probability of survival of different droplets and compare it with some estimate connected to the {\it exit-time} along the droplet trajectory \citep{Baebler12}, i.e. the average time it takes for a droplet to experience a total stress strong enough to break it. This is connected to the Lagrangian time-decorrelation, persistency and efficiency of stress along the trajectory. A study in this direction is left for future work. 
For cases when the viscosities of droplet and outer fluid differ, such that 
vorticity is able to decorrelate the droplet   from the straining directions, the large deviation theory prediction fails.  
The results highlight the complex dynamics of droplet deformation and 
orientation and opens the way to estimate/model the feedback on the flow due to the presence of deformable droplets. 

The case of droplets/bubbles with a large density mismatch with respect to the density of the 
 underlying fluid can be treated with the same approach of M\&M  to study the 
deformation along point-like particles but following inertial trajectories \citep{Becetal-jfm-2010} instead of fluid tracers as done here. For situations in which there is a slip velocity between the droplets and fluid, it is necessary to add the stress induced by the  Stokes drag in order to evaluate the deformation of the droplet. 
\section*{Acknowledgements}
The authors are indebted to Prof. Mario Minale for having provided some relevant literature and useful advice. L.B.  acknowledges partial funding from the European Research Council under the European Community's Seventh Framework Program,  ERC Grant Agreement No 339032. C.M. is grateful to the Universit\'{a} di Roma ``Tor Vergata''  for their hospitality and support during a one month visit in May 2013, and to GoMRI for partial funding of research on droplets in turbulence. 

\bibliographystyle{jfm}
\bibliography{turbulence}

\end{document}